\documentclass[prl,twocolumn]{revtex4}
\usepackage{graphicx}
\renewcommand{\v}[1]{{\bf #1}}

\newcommand{\s}{{\sigma}}

\newcommand{\w}{{\omega}}

\def\eqa{\begin{eqnarray}}
\def\eea{\end{eqnarray}}
\newcommand{\eq}{\begin{equation}}
\newcommand{\ee}{\end{equation}}

\newcommand{\<}{\langle}
\renewcommand{\>}{\rangle}

\renewcommand{\Im}{{\rm Im}}

\newcommand{\ua}{\uparrow}
\newcommand{\da}{\downarrow}
\newcommand{\ra}{\rightarrow}

\newcommand{\al}{\alpha}

\newcommand{\del}{\delta}
\newcommand{\Del}{\Delta}

\newcommand{\ga}{\gamma}
\newcommand{\Ga}{\Gamma}

\newcommand{\si}{\sigma}
\newcommand{\Si}{\Sigma}

\newcommand{\cN}{ {\cal N} }

\begin{document}

\title{Electronic Raman scattering in superconductors on triangle lattices}
\author{Jian-Bo Wang, Xiang-Mei He, Fei Tan, and Qiang-Hua Wang}
\affiliation{National Laboratory of Solid State Microstructures,
Nanjing University, Nanjing 210093, China}


\begin{abstract}
In superconductors on triangle lattices, we show that the
electronic Raman spectra in the non-resonant channel are
independent of the light polarization configuration in the
scattering. Therefore the light polarization configuration is no
longer a judicious choice to probe the anisotropic nature of the
pairing wave function (as in the case of square lattices).
However, we show that the detailed energy-shift dependence of the
spectrum and the effects of impurities provide useful information
that can help identify the possible pairing symmetry
experimentally. These results are relevant to the newly discovered
Na$_x$CoO$_2\cdot y$H$_2$O superconductors.
\end{abstract}

\pacs{PACS numbers: 74.20.Rp, 74.25.Nf, 74.20.-z} \maketitle

The electronic Raman scattering in the so-called non-resonant
channel (without invoking inter-band scattering) measures the
fluctuations of the effective density operator $\rho_{\v
q}=\sum_{\v k\s} \ga_{\v k} C^\dagger_{\v k+\v q\s}C_{\v k\s}$ in
the limit of $\v q\ra 0$ (In the scattering event, $\v q$ is the
momentum transfer from the lights, which is vanishingly small
compared to the Fermi momentum.) Here $\ga_{\v k}$ is the Raman
vertex that describes the diamagnetic second-order coupling
between electrons and lights, $\ga_\v k=(\v e_i\cdot \nabla_{\v
k})(\v e_s\cdot \nabla_{\v k})\epsilon_{\bf k}$, where $\v e_i$
and $\v e_s$ are the polarization directions of the incident and
scattered lights, and $\epsilon_{\v k}$ is the normal state
dispersion. The Raman response function $\chi(\v q,\w)$, with
$\w=\w_i-\w_s$ being the energy transfer in the scattering, is the
Fourier transform of the retarded Green's function $\chi(\v
q,t)=-i\< [\rho_{\v q}(t),\rho_{-\v q}(0)]\>\theta(t)$. The Raman
intensity ($RI$) is eventually given by $-\Im[\chi(\v q\ra
0,\w)]$. Theoretically, $\chi(\v q,\w)$ can be conveniently
obtained from the analytical continuation of the Matsubara
propagator $\chi(\v q,i\w_n\ra \w+i0^+)$.

The electronic Raman response vanishes in ideal metals because of
lack of phase space for finite-energy and zero-momentum
particle-hole excitations, but is finite in a superconductor
because a Cooper pair can be broken with no change in the
center-of-mass momentum. By adjusting the incident and scattered
photon polarizations, through the Raman vertex function Raman
scattering may projects out various parts of the quasi-particle
excitations in the momentum space, and is therefore highly useful
to probe the angular dependence of the possible anisotropic
pairing wave function on the Fermi surface. This has been applied
very successfully in cuprate superconductors.\cite{hightcraman}
Because of the square lattice of the copper oxide plane, there are
two Raman-active anisotropic channels, {\it i.e.}, the $B_{1g}$
and $B_{2g}$ channels. These channels can be conveniently picked
up by adjusting the polarizations of the incident and scattered
lights. The fact that the Raman responses are different in these
channels is a proof that the pairing wave function is anisotropic,
and is consistent with the $d$-wave symmetry. Furthermore, the
detailed energy-dependence of the response yield useful
information on the dynamics of the quasi-particles. It is
therefore natural to explore how electronic Raman scattering can
be used in superconductors on other lattices.

In this paper, we shall concentrate on triangle lattices,
motivated by the discovery of the Na$_x$CoO$_2\cdot y$H$_2$O
superconductor,\cite{takada} which does have triangular CoO$_2$
layers, separated by thick insulating layers of Na$^+$ ions and
H$_2$O molecules. The two-dimensional layered structure is similar
to the copper oxide and it is conjectured that a similar
superconductivity mechanism may be at work. In particular, it is
argued that this is a doped Mott insulator that would be described
by the resonant-valence-bond (RVB) theory.\cite{rvb} In such a
scenario, the pairing symmetry would be $d_{x^2-y^2}+id_{xy}$ with
a full energy gap for quasi-particle excitations. On the other
hand, given the experimental fact that the parent compounds
exhibit intra-plane ferromagnetic correlations, $p_x+ip_y$-
\cite{pwave} or $f$-wave \cite{fwave} pairing symmetries are also
proposed as possible candidates. For these triplet channels, the
$p$-wave pairing has a full gap, while the $f$-wave pairing have
nodal lines in the gap function. Finally, if the pairing is due to
lattice vibrations, an $s$-wave gap function is to be expected. As
an important step toward understanding the pairing mechanism, one
is therefore asked to discriminate experimentally one from the
others in this list of possible pairing symmetries.

Existing experimental data are still controversial. From
nuclear-magnetic-resonance (NMR) measurements, Kato {\it et al}
proposed $p$-wave triplet pairing,\cite{kato} while Kobayashi {\it
et al} conjectured $s$-wave singlet pairing.\cite{kobayashi}
Nonetheless, from the coherence peak in the temperature dependence
of the spin-lattice relaxation rate $1/T_1$ both groups agreed
that the gap has no nodal lines. This is to be contrasted to the
evidence of nodal lines in nuclear-quadruple-resonance (NQR) data
\cite{nqr} and in the specific heat measurement.\cite{heat} The
main difficulty for all measurements is the control of the sample
quality, and it is quite likely the reason behind the
inconsistency of the experimental results. While this asks for
better sample quality to resolve the inconsistency, in this paper
we propose electronic Raman scattering as additional but important
probe to the pairing symmetry. We note, however, that the low
transition temperature (about $4.5$K) may impose a difficulty for
the measurement of the pairing gap due to the laser heating
effect. Our hope is that the technical difficulty could be removed
and that samples with higher transition temperatures could be
achieved.

The structure of the rest of the paper is as follows. We first
analyze the Raman response in the clean system, reaching the
result that the response is independent of the polarization
configuration of the scattering. We show the result by explicit
calculation of a lattice model, and substantiate by a proof in the
continuum limit. The Raman response in the clean system can only
tell whether there is a full gap on the Fermi surface. Second, the
effect of impurity is taken into account. We find that the
impurity effects do not induce Raman absorption below the pair
breaking peak (PBP) for $s$-wave pairing, while they do so for
$p_x+ip_y$- and $d_x^2-y^2+id_{xy}$-wave pairing. We also find a
universal Raman response in the case of nodal $f$-wave pairing in
the limit of small Raman energy shift for small scattering rates.
Combined with a probe to the singlet versus triplet pairing, these
features enables one to identify the pairing symmetry
unambiguously.

We start from the effective mean field Hamiltonian,\eqa H=\sum_{\v
k\s}\epsilon_{\v k}C_{\v k\s}^\dagger C_{\v k\s}+\sum_{\v k}
(\Del_{\v k}C_{\v k\ua}^\dagger C_{-\v k\da}^\dagger +{\rm
h.c.}),\eea where $\epsilon_{\v k}=-2t\sum_{i=1}^3\cos \v k\cdot
\v \del_i-\mu$. Here $\v \del_{1,2,3}=(1,0),\ (1/2,\sqrt{3}/2)$
and $(-1/2,\sqrt{3}/2)$ are three unit vectors on the triangle
lattice, $\mu$ is the chemical potential. For the
Na$_x$CoO$_2\cdot y$H$_2$O superconductor, band structure
calculation reveals a large hole pocket around the $\Ga$ point in
the momentum space,\cite{singh} it is therefore consistent to take
$t<0$ in the dispersion. We also tune the chemical potential so
that there are $1.35$ electrons per site. $\Del_{\v k}$ is the
pairing gap function. For the $s$-, $p_x+ip_y$-,
$d_{x^2-y^2}+id_{xy}$- and $f$-wave pairing symmetries, the gap
functions we will consider are, respectively, \eqa
\Del_\v k^s=\Del \sum_{i=1}^3 \cos \v k\cdot\v\del_i;\\
\Del_\v k^p=\Del \sum_{i=1}^3 \sin (\v
k\cdot\v\del_i)\exp(i\theta_i);\\
\Del_\v k^d=\Del\sum_{i=1}^3\cos(\v k\cdot\del_i)\exp(2i\theta_i);\\
\Del_\v k^f=\Del\sum_{i=1}^3\sin(\v k\cdot\del_i)\exp(3i\theta_i),
\eea where $\theta_{1,2,3}=0,\pi/3$ and $2\pi/3$ are the azimuthal
angles of $\v\del_{1,2,3}$, respectively. In the following
calculation we take the gap amplitude as $\Del=0.1|t|$ for
illustration purpose. The conclusion does not depend on this
parameterization. We note that the gap on the Fermi surface is
finite for the cases of $s$-, $p_x+ip_y$-, and
$d_{x^2-y^2}+id_{xy}$-wave cases, and it has nodal points for the
$f$-wave pairing. The bare Matsubara propagator for the effective
density $\rho_\v q$ in the limit of $\v q=0$ can be written as,
\eqa & &\chi_{\ga\ga}^0(i\nu_n)=\nonumber\\ & &
-\frac{T}{N}\sum_{\v k,i\w_n}{\rm Tr} [\ga_\v k\s_3G(\v
k,i\w_n)\ga_\v k\s_3G(\v k,i\w_n+i\nu_n)], \label{chi0}\eea where
$N$ is the number of lattice sites, and $G$ is the mean field
Green's function given by \eqa G^{-1}(\v
k,i\w_n)=i\w_n\s_0-\epsilon_\v k\s_3-(\Del_\v k\s^++{\rm
h.c.}).\eea Here $\w_n$ and $\nu_n$ are Fermionic and Bosonic
Matsubara frequencies, respectively, $\s_0$ is the $2\times 2$
unitary matrix, $\s_{1,2,3}$ are the Pauli matrices, and
$\s^+=(\s_1+i\s_2)/2$. The summation over Matsubara frequencies
can be completed exactly. After the analytical continuation, we
get the bare retarded Raman response function, \eqa
\chi_{\ga\ga}^0(\w)=\frac{1}{N}\sum_\v k \frac{\ga_\v
k^2|\Del_\v k|^2}{E_\v k^2}\tanh \frac{\beta E_\v k}{2}\nonumber\\
\times [\frac{1}{\w-2E_\v k+i0^+}-\frac{1}{\w+2E_\v k+i0^+}],\eea
where $E_\v k=\sqrt{\epsilon_\v k^2+|\Del_\v k|^2}$. In the limit
where $\ga_\v k\sim 1$ the response function is equivalently a
bare charge density susceptibility in the long wave length limit.
However, in a charged system, the spectral weight of this
susceptibility is pushed up to the plasma frequency. While this
fact can be taken into account by a
random-phase-approximation-like approach, the effect of Coulomb
screening on the Raman resposne can be taken into account by
simply projecting away the symmetrical part as
follows\cite{screening} \eqa \chi_{\ga\ga}=
\chi_{\ga\ga}^0-\frac{\chi_{\ga
1}^0\chi_{1\ga}^0}{\chi_{11}^0},\label{chi}\eea which is
applicable below the plasma frequency. Here we suppressed the
frequency arguments for brevity, and $\chi_{\ga 1}^0$ and
$\chi_{11}^0$ are obtained by replacing $\ga_\v k$ by $1$ for one
or two of the Raman vertices in $\chi_{\ga\ga}^0$.

To proceed, we write according to the given definition the three
typical Raman vertex functions in the $xx$, $yy$ and $xy$
polarization configurations as follows, \eqa \ga_\v k^{xx}=2t(
\cos k_x+\frac{1}{2}\cos\frac{k_x}{2}\cos\frac{\sqrt{3}k_y}{2});\\
\ga_\v k^{yy}=3t\cos\frac{k_x}{2}\cos\frac{\sqrt{3}k_y}{2};\\
\ga_\v
k^{xy}=-\sqrt{3}t\sin\frac{k_x}{2}\sin\frac{\sqrt{3}k_y}{2}.\eea
The vertex function for an arbitrary polarization configuration
can be expressed in terms of the above as,\eqa \ga_\v
k=\cos\phi_i\cos\phi_s\ga_\v
k^{xx}+\sin\phi_i\sin\phi_s\ga_\v k^{yy}\nonumber\\
+\sin(\phi_i+\phi_s)\ga_\v k^{xy},\eea where $\phi_{i,s}$ are the
azimuthal angles of the polarizations of the incident and
scattered lights, respectively.

With the above ingredients, we are now able to calculate the Raman
spectra (the imaginary parts of $\chi_{\ga\ga}^0$ and
$\chi_{\ga\ga}$) explicitly. We take the zero temperature limit in
what follows. Fig.1 is the spectra for the $p_x+ip_y$-wave pairing
in the $xx$ (a), $yy$ (b) and $xy$ (c) polarization
configurations. The dashed (solid) lines denote the bare
(screened) spectra. The PBP occurs near $\w=4\Del$. This is
because the effective gap on the Fermi surface is near $2\Del$,
and the major contribution to the summation over $\v k$ in
$\chi^0$ is from the region near the Fermi surface. It is clear
that while the bare spectra depends on the polarization
configurations, the screened ones do not, which is made clearer by
plotting them together in Fig.1(d). This is true for any other
arbitrary configurations (not shown here). Experimentally it is
the screened spectra that are measured. We therefore conclude that
the Raman spectra on triangle lattices are independent of the
polarization configuration.

\begin{figure}
\includegraphics[width=8.5cm]{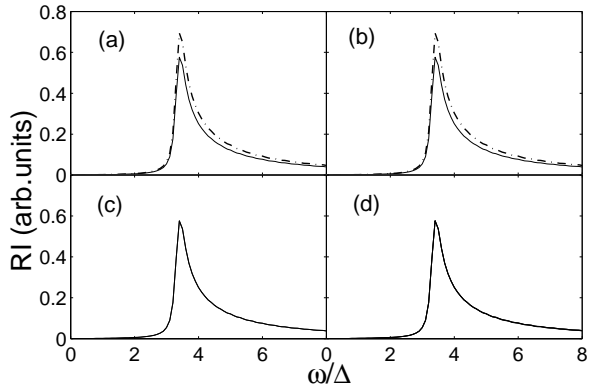}
\caption{The electronic Raman spectra for $p_x+ip_y$-wave pairing
in the (a) $xx$, (b) $yy$ and (c) $xy$ polarization
configurations. The dashed (solid) lines denote the bare
(screened) spectra. (d) is the screened spectra in the three
configurations plot together. Note that in (c) the dashed line
falls on top of the solid line and is therefore invisible.}
\end{figure}

In order to understand the above result, it is useful to resort to
symmetry analysis. In general, the Raman vertex function can be
decomposed in terms of the basis function of the irreducible
representations of the point group. For the system under concern,
there are only two Raman-active irreducible representations, {\it
i.e.}, the one-dimensional $A_{1g}$ and two-dimensional $E_g$
representations. The corresponding basis function for $A_{1g}$ is
\eqa \ga_a=\cos
k_x+2\cos\frac{k_x}{2}\cos\frac{\sqrt{3}k_y}{2},\eea and that for
the $E_g$ representation is $\ga_{e}=(\ga_{e1},\ga_{e2})$ with
\eqa
\ga_{e1}=\sqrt{3}\sin\frac{k_x}{2}\sin\frac{\sqrt{3}k_y}{2};\\
\ga_{e2}=\cos k_x-\cos\frac{k_x}{2}\cos\frac{\sqrt{3}k_y}{2}.\eea
Symmetry arguments guarantee that the Raman response is also
decomposed into these two channels. Furthermore, since $\ga_a$ has
the full symmetry of the system itself, the Raman response in this
channel will be screened. Therefore, $E_g$ is the only channel for
electronic Raman response, and this implies that at least the line
shape of the Raman response should be independent of the
polarization configuration as illustrated in Fig.1. It remains to
be shown that even the amplitude of the Raman spectra is
independent of the polarization configuration. We first verify the
claim by taking advantage of the typical result in Fig.1, and then
present a more rigorous proof by going to the continuum limit.

In terms of the basis functions, the three typical Raman vertex
functions can be rewritten as, \eqa
\ga^{xx}=t(\ga_a+\ga_{e2});\\
\ga^{yy}=t(\ga_a-\ga_{e2});\\
\ga^{xy}=-t\ga_{e1}.\eea Since the $\ga_a$ component does not
contribute to $\chi_{\ga\ga}$ due to screening, Fig.1 proves that
the $\ga_{e1,e2}$ components contribute equally to the
$\chi_{\ga\ga}$, or
$\chi_{\ga_{e1}\ga_{e1}}=\chi_{\ga_{e2}\ga_{e2}}$. Now consider an
arbitrary vertex which can be rewritten as \eqa
\ga=t(\cos\phi\ga_{e1}-\sin\phi\ga_{e2}),\eea where
$\phi=\phi_i+\phi_f$, and we dropped the $\ga_a$ component,
anticipating the screening effect. Consequently, \eqa
\chi_{\ga\ga}=\cos^2\phi
\chi_{\ga_{e1}\ga_{e1}}+\sin^2\phi\chi_{\ga_{e2}\ga_{e2}}\equiv
\chi_{\ga_{e1}\ga_{e1}},\eea where a contribution from the
crossing term $\ga_{e1}\ga_{e2}$ vanishes by symmetry.

To close the argument, we now prove
$\chi_{\ga_{e1}\ga_{e1}}=\chi_{\ga_{e2}\ga_{e2}}$ for all of the
pairing symmetries listed previously in the continuum limit. Since
the screening effect is absent in the anisotropic $E_g$ channel,
it suffices to prove
$\chi^0_{\ga_{e1}\ga_{e1}}=\chi^0_{\ga_{e2}\ga_{e2}}$. Near the
Fermi surface, one can write approximately but without loss of
symmetry (up to trivial common factors), \eqa \ga_{e1}\sim \sin
2\theta; \ \ga_{e2}\sim \cos 2\theta,\eea where $\theta$ is the
azimuthal angle of the Fermi momentum $\v k_F$ approximately on a
circle. On the other hand, the gap functions
can be approximated as \eqa \Del_\theta^s\sim \Del_F,\\
\Del_\theta^p\sim \Del_F \exp(i\theta),\\ \Del_\theta^d\sim
\Del_F\exp(2i\theta),\\ \Del_\theta^f\sim \Del_F\sin(3\theta),\eea
where $\Del_F$ is the effective gap amplitude on the Fermi surface
for the specific cases. It is seen that $|\Del_\theta|^2$ is
either a constant (for $s$-, $p_x+ip_y$- and
$d_{x^2-y^2}+id_{xy}$-symmetries) or six-fold symmetric (for
$f$-wave symmetry) on the Fermi surface, whereas
$\ga_{e1,e2}^2=(1\pm \cos 4\theta)/2$ which is four-fold
symmetric. Plugging these into the expression for $\chi^0$ and
going over to the continuum limit, one concludes that by symmetry
only the constant part in $\ga_{e1,e2}^2$ contributes to the
response function. This complete the prove. It is interesting to
point out that in the case of $d_{x^2-y^2}$-wave pairing on a
square lattice, $\ga_{e1,e2}$ becomes the basis functions for the
$B_{2g}$ and $B_{1g}$ representations. In this case,
$|\Del_\theta|^2$ is four-fold symmetric, which can resonate with
the corresponding component in $\ga_{e1,e2}^2$, and this is why in
that case the Raman spectra is different in $B_{1g}$ and $B_{2g}$
channels.

Having convinced ourselves that the Raman spectra on a triangle
lattice do not depend on polarization configuration for all the
relevant pairing symmetries, we now look into the energy
dependence of the spectra. Fig.2 shows the screened Raman spectra
for the four pairing symmetries under concern. As anticipated, the
spectra have a threshold for the fully-gapped (a) $s$-, (b)
$p_x+ip_y$-, and (c) $d_{x^2-y^2}+id_{xy}$-wave pairing, whereas
it is linear at low energy for the nodal (d) $f$-wave pairing. The
PBP in Fig.2 occur at different energies in units of $\Del$. This
is again due to the fact that the effective gap amplitude $\Del_F$
on the Fermi surface differs in the cases considered, even though
the same $\Del$ is used in the gap function. We also note that the
line shape for the $d$-wave case is slightly different from the
$s$- and $p$-wave cases. This is because the $d$-wave gap value
under concern is close to the van-Hove singularity of the normal
state density of states (DOS), which contributes a secondary peak
near the threshold. We do not expect it to be a generic feature.
In summary, in the clean system considered so far, the Raman
spectra can only tell whether the pairing gap is nodal or not.
However, impurities are always present in real systems, and it is
pending to see whether impurity effects can help differentiate the
three fully-gapped pairing cases. This is to be anticipated as
impurities may induce low energy (Andreev) states for nontrivial
pairing, even if it is fully gapped in the clean system. This will
be discussed in the following.

\begin{figure}
\includegraphics[width=8.5cm]{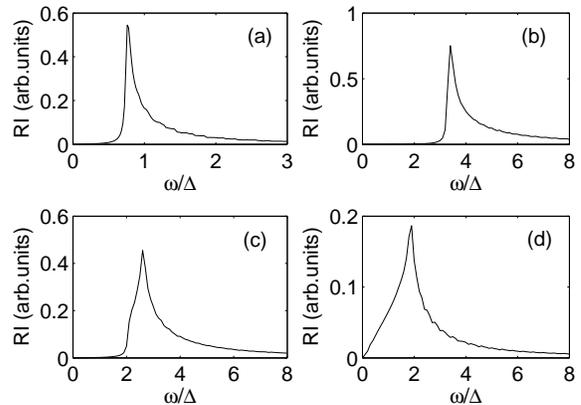}
\caption{The screened electronic Raman spectra for (a) $s$-, (b)
$p_x+ip_y$-, (c) $d_{x^2-y^2}+id_{xy}$-, and (d) $f$-wave
pairing.}
\end{figure}

The effect of impurities can be conveniently accounted for by the
coherent potential approximation (CPA). The modified Green's
function is determined by \eqa G^{-1}(\v
k,i\w_n)=i\w_n\s_0-\epsilon_\v k\s_3-(\Del_\v k\s^++{\rm
h.c.)-\Si(\w_n},\eea where
$\Si(\w_n)=\Si_0(\w_n)\s_0+\Si_1(\w_n)\s_1$. Henceforth we apply
the continuum limit and assume particle-hole symmetry near the
Fermi surface so that the $\s_3$ component of $\Si$ vanishes. We
take $\Del_\v k^s$ to be real for definiteness. On the other hand,
we anticipate that the $\s_{1,2}$ (or $\si^{\pm}$) components of
$\Si$ vanishes for pairing symmetries other than the $s$-wave, as
to be verified hereafter. The self-energy $\Si$ is determined by
the self-consistent condition \eqa \Si=-\frac{\Ga}{g-c\s_3
g^{-1}\s_3},\eea where $\Ga=n_i/\pi\cN_0$ is the scattering rate
with $n_i$ and $\cN_0$ being the density of impurities and the DOS
at the Fermi energy, respectively, $c=\cot\delta$ with $\delta$
being the scattering phase shift, and $g$ is the normalized
on-site Green's function \eqa g=\frac{1}{\pi\cN_0}\int\frac{d^2\v
k}{(2\pi)^2}G(\v k,i\w_n).\eea It is straightforward to verify
that the $\si_3$ component of $g$ is absent by particle-hole
symmetry, and so are the $\si_{1,2}$ components in the case of
non-$s$-wave pairing, justifying the {\it a priori} assumptions.
Similar self-consistent equations holds for the retarded and
advanced Green's functions, and are used in the realistic
calculations via the spectral representation,\eqa G(\v
k,i\w_n)=\frac{1}{2\pi i}\int \frac{G^A(\v k,\w)-G^R(\v
k,\w)}{i\w_n-\w}d\w.\label{spectralg}\eea This is substituted into
Eqs.(\ref{chi0}) and (\ref{chi}) to calculate the Raman spectra in
the presence of impurities. \cite{cparaman} Vertex corrections
have been ignored, and are not expected to affect the conclusions
significantly.

We first discuss the effect of unitary limit ($c=0$). Fig.3(a) and
(c) show the Raman spectra for $s$- and $p_x+ip_y$-wave pairing
under various values of the scattering rate $\Ga$. For the
$s$-wave pairing, the Raman spectra remains forbidden below the
PBP at $\w=2\Del_F$. Impurity scattering merely changes slightly
the spectra above the peak. This is easily understood from the
behavior of the DOS plot in Fig.3(b), where impurity does not
induce intra-gap quasi-particles, consistent with the Anderson
theorem. The slight change in the Raman spectra is due to the
slight change in the DOS above the gap energy. The situation is
quite different in Fig.3(c), where the Raman spectra is induced
below the PBP. This can again be understood from the DOS induced
by impurities in Fig.3(d). While there is a full gap in the clean
limit, an impurity band is induced near the Fermi energy for small
$\Ga$. This is due to the nontrivial intrinsic phase of the
$p_x+ip_y$-wave pairing function. During impurity scattering, the
quasi-particles experience a change of the intrinsic phase. It is
shown elsewhere that by such an effect a single unitary impurity
can induce an Andreev bound state, \cite{impurityboundstate}and
the bound states from a distribution of impurities can therefore
form an impurity band. The scattering of quasi-particles by the
light from the impurity band (below the Fermi energy) to the gap
edge (above the Fermi energy) contributes most significantly to
the Raman absorption below the PBP. This also explains the hump
with a pseudo-threshold at $\w=\Delta_F$. For very large $\Ga$,
the intra-gap is filled with induced DOS and accordingly the Raman
spectra varies smoothly below $2\Del_F$. We add that the Raman
spectra for $d_{x^2-y^2}+id_{xy}$-wave pairing behaves similarly
to the $p_x+ip_y$-wave pairing under a similar mechanism, and are
not shown here for brevity.

\begin{figure}
\includegraphics[width=8.6cm,height=6.5cm]{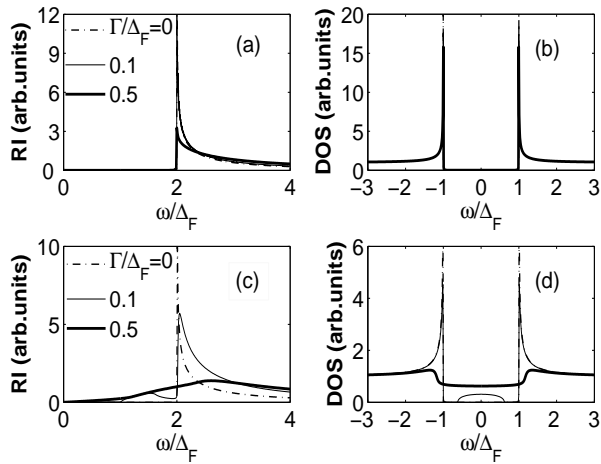}
\caption{The electronic Raman spectra for (a) $s$- and (c)
$p_x+ip_y$-wave pairing in the unitary limit for various values of
the scattering rate. The corresponding DOS are shown for (b) $s$-
and (d) $p_x+ip_y$-wave pairing.}
\end{figure}

Fig.4(a) shows the Raman spectra in the unitary limit for $f$-wave
pairing. Although the PBP is lowered with increasing scattering
rate, the low energy part is hardly changed. This implies a
universal low energy Raman spectra in this case, to which we shall
return shortly. The DOS is plot in Fig.4(b). In the absence of
impurities, the DOS is linear at low energy, and impurities again
induce low energy states. However, the impurity band is different
from that in the $p_x+ip_y$-wave case, in that this band connects
smoothly to the higher energy DOS. This explains the absence of
new humps below the PBP in the Raman spectra. The universal low
energy Raman spectra deserves further scrutiny. For a low
scattering rate $\Ga$, this can be proved exactly. From the CPA
self-consistent equation the retarded self-energy $\Si^{A,R}=\pm
i\eta\si_0$ in the low energy limit, where $\eta\sim
\sqrt{\Ga\Del_F}$. If this is substituted into
Eqs.(\ref{spectralg}) and (\ref{chi0}), we find after some algebra
that for $\w\ll \Del_F$ the Raman spectra is given by $-\Im
\chi(i\nu_n\ra \w+i0^+)=\al\w\ga^2/v_F v_\Del$, where $\al$ is a
constant, $\ga^2=\<\ga_\v k^2\>_F$ is the average square of the
Raman vertex on the Fermi surface, $v_F$ is the Fermi velocity and
$v_\Del=|\nabla_\v k\Delta_\v k|_{\v k=\v k_n}$ is the gap
velocity at the six nodal points $\v k_n$ on the Fermi surface.
The result is tied to the existence of nodal quasi-particles. The
universal behavior in the Raman spectra is reminiscent to the
universal conductivity discovered in the literature for
$d_{x^2-y^2}$-wave pairing in the cuprate
superconductors.\cite{universal}

\begin{figure}
\includegraphics[width=8.5cm]{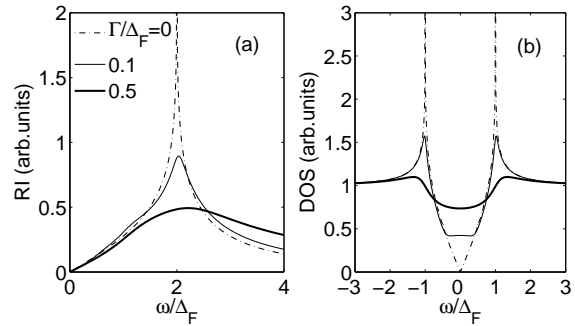}
\caption{(a) The electronic Raman spectra for $f$-wave pairing in
the unitary limit. (b) The corresponding DOS. }
\end{figure}

We have also considered general values of the inverse scattering
strength $c$. The conclusion does not change for the $s$-wave
case. The Raman spectra for $p_x+ip_y$-wave case is shown in
Fig.5(a) for various values of $c$. Fig.5(b) is the corresponding
DOS. It is seen that near the Born limit, the center of the
impurity band shifts away from the Fermi energy, causing a
corresponding change in the Raman spectra. The Raman spectra for
$f$-wave case is shown in Fig.5(c), and the DOS in Fig.5(d). Here
no impurity band is formed in the Born limit, and the zero-energy
DOS increases gradually with decreasing $c$. Interestingly this
does not cause a change in the low energy limit of the Raman
spectra. The reason is that for any values of $c$, as long as the
resulting self energy is independent of energy in the low energy
limit, the mechanism of universal Raman spectra applies as
described above.

\begin{figure}
\includegraphics[width=8.5cm,height=6.5cm]{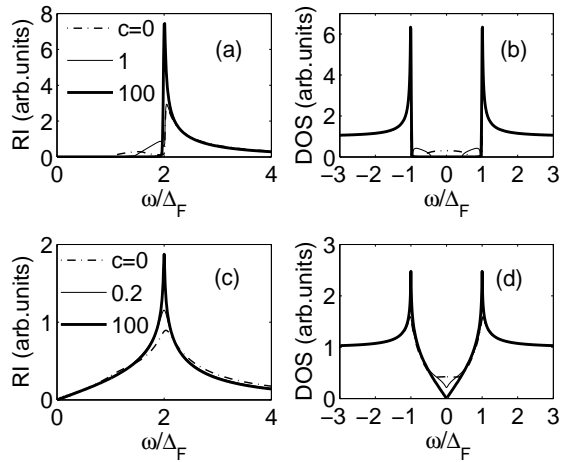}
\caption{The electronic Raman spectra for (a) $p_x+ip_y$- and (c)
$f$-wave pairing with $\Ga/\Delta_F=0.1$ and various values of
$c$. The corresponding DOS are shown for (b) $p_x+ip_y$- and (d)
$f$-wave pairing.}
\end{figure}

In summary, we developed a theory of electronic Raman scattering
in superconductors on triangle lattices, with possible application
to the cobalt oxide superconductors. The Raman response is shown
to be independent of the light polarization configuration.
However, in the case of $s$-wave pairing the absorption is
forbidden below the PBP in the clean system as well as in the
presence of impurity scattering; in the cases of $p_x+ip_y$- and
$d_{x^2-y^2}+id_{xy}$-wave pairing the absorption below the PBP is
forbidden in the clean system, but can be induced by impurities;
and finally in the case of nodal $f$-wave pairing there exists a
universal low energy Raman response in the low energy limit for
weak scattering rates. These features, combined with another probe
of singlet versus triplet pairing, can be used to identify the
relevant pairing symmetries unambiguously.

\acknowledgements{We thank Qing-Ming Zhang for useful discussions.
This work was supported by NSFC 10204011, 10429401 and 10021001,
the Fok Ying Tung Education Foundation No.91009, and the Ministry
of Science and Technology of China (973 project No:
2006CB601002).}

\end{document}